\documentclass[twocolumn,aps,prb,amsmath,floatfix]{revtex4}
\usepackage{graphicx}
\begin{document}
\title{Spectral weight of doping-induced states in the 2D Hubbard model}
\author{Ansgar Liebsch}
\affiliation{Institut f\"ur Festk\"orperforschung, 
             Forschungszentrum J\"ulich, 
             52425 J\"ulich, Germany} 
\begin{abstract}
The spectral weight of states induced in the Mott gap via hole doping 
in the two-dimensional Hubbard model is studied within cluster dynamical 
mean field theory combined with finite-temperature exact diagonalization.
If the cutoff energy is chosen to lie just below the upper Hubbard band, 
the integrated weight per spin is shown to satisfy $W_+(\delta)\ge\delta$ 
($\delta$ denotes the total number of holes), in           agreement with 
model predictions by Eskes {\it et al.} [Phys. Rev. Lett. {\bf 67}, 1035
(1991)]. However, if the cutoff energy is chosen to lie in 
the range of the pseudogap, $W_+(\delta)$ remains much smaller than 
$\delta$ and approximately saturates near $\delta\approx 0.2\ldots0.3$.
The analysis of recent X-ray absorption spectroscopy data therefore 
depends crucially on the appropriate definition of the integration window.  
\\
\mbox{\hskip1cm}  \\
PACS. 71.20.Be  Transition metals and alloys - 71.27+a Strongly correlated
electron systems 
\end{abstract}
\maketitle

\section{Introduction}

The two-dimensional single-band Hubbard model has been widely used to 
study the role of Coulomb correlations in the high-$T_c$ cuprates.
One of the remarkable features of this model was pointed out long ago
by Eskes {\it et al.} \cite{eskes}, namely, that doping the system
with $\delta$ holes does not yield unoccupied low-energy states of weight
$\delta/2$ per spin, like in an ordinary band insulator. Instead, 
as a result of strong local Coulomb interactions,
this weight is approximately given by $W_+(\delta)\ge\delta$. 
The physical reason for this feature is that both lower and upper 
Hubbard bands must contribute to the generation of itinerant low-energy 
states when holes are added to the system.

In striking contrast to this prediction, recent  X-ray absorption 
spectroscopy (XAS) data on Tl$_2$Ba$_2$CuO$_{6-\delta}$ and
La$_{2-x}$Sr$_x$CuO$_{4\pm\delta}$ by Peets {\it et al.} \cite{peets} 
show a linear 
behavior $W_+(\delta)\approx\delta$ only up to about $\delta\approx 0.2$.
At doping concentrations in the range $\delta=0.2\ldots0.3$,  $W_+(\delta)$  
levels off, suggesting the inapplicability of the single-band Hubbard 
model for these high-$T_c$ compounds. 

Since the calculations by Eskes {\it et al.} \cite{eskes} were carried 
out for small one-dimensional clusters, it is not entirely clear 
to what extent the discrepancies with respect to the data in Ref.\cite{peets} 
might be related to the simplicity of the theoretical model. 
In fact, Phillips and Jarrell\cite{phillips} recently claimed
that state-of-the-art many-body calculations based on the dynamical
cluster approximation (DCA)\cite{hettler,jarrell2001} do indeed predict a
saturation of $W_+(\delta)$ close to $\delta\approx 0.2$, in agreement
with the measurements by Peets {\it et al.}                

The aim of this work is to demonstrate that the doping variation 
of the spectral weight of the induced low-energy  states depends crucially 
on the choice of the upper limit of the energy window in which these 
states are counted.\cite{peets2}
To evaluate $W_+(\delta)$  for the two-dimensional Hubbard
model we use the cluster extension of dynamical mean field theory (DMFT)
\cite{dmft,kotliar01} combined with finite-temperature exact diagonalization
(ED).\cite{prb09} The results show that, if the integration window of 
$W_+(\delta)$ is chosen to reach up to the lower edge of the upper Hubbard 
band, then $W_+(\delta)\ge\delta$, just as predicted by Eskes {\it et al.}\cite{eskes}  
If the cutoff energy, however, is chosen to lie in the range of the 
pseudogap, then $W_+(\delta)$ is much smaller than $\delta$ and approximately 
saturates near $\delta\approx 0.2\ldots0.3$, as found in Ref.\cite{phillips}
 It is clear, therefore, that
the interpretation of the XAS measurements must be based on the correct
choice of the energy window over which the induced low-energy states are 
taken into account.

\section{Results and discussion}

The cluster DMFT calculations are carried out for the two-dimensional
Hubbard model with nearest and next-nearest hopping parameters $t=0.25$~eV
and $t'=-0.075$~eV, respectively (band width $W=2$~eV).
The onsite Coulomb interaction is $U=2.5$~eV and the temperature is $T=0.01$~eV. 
With this choice of parameters the system is a Mott insulator in the zero
doping limit. To account for intersite correlations, the square lattice is 
viewed as a superlattice consisting  of $2\times2$ clusters. 
Details of these finite-temperature ED / DMFT calculations can be found 
in Ref.\cite{prb09}

\begin{figure} 
\begin{center}
\includegraphics[width=10.0cm,height=6.5cm,angle=-90]{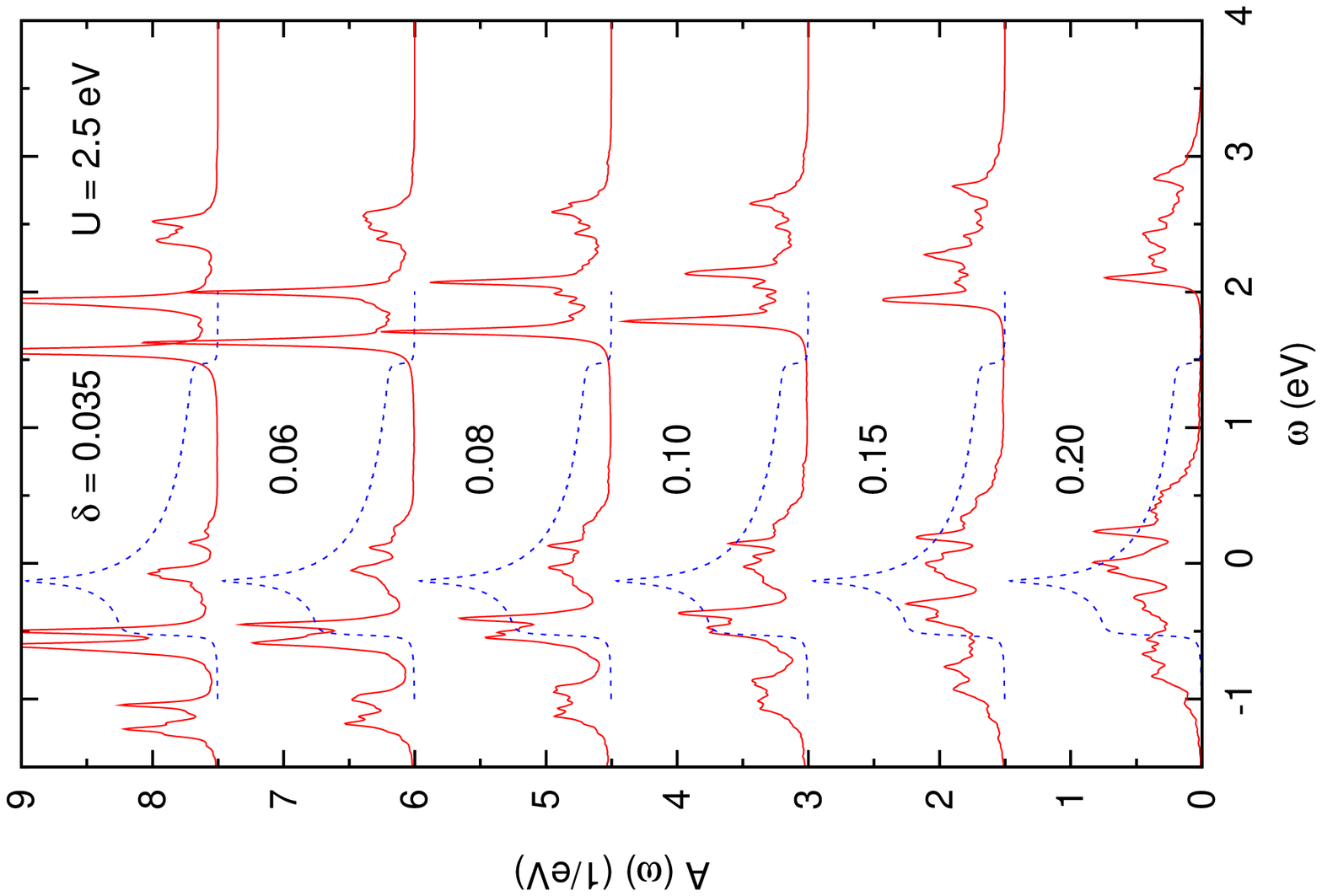}
\end{center}
\vskip-7mm \ \ \ (a)\hfill  \mbox{\hskip5mm}
\begin{center}
\vskip-5mm
\includegraphics[width=10.0cm,height=6.5cm,angle=-90]{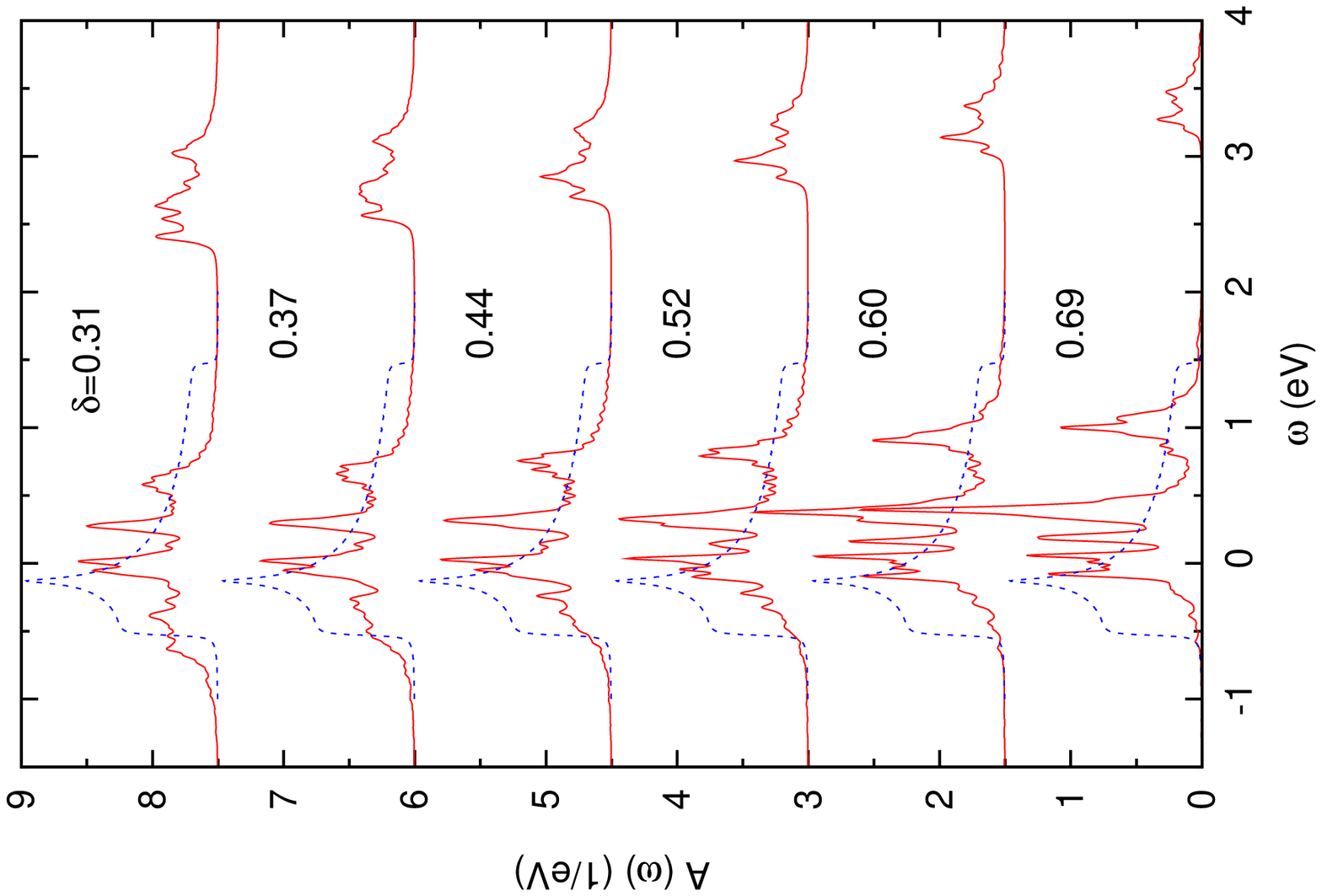}
\end{center}
\vskip-7mm \ \ \ (b)\hfill  \mbox{\hskip5mm}
\caption{(Color online)
Spectral distributions calculated within cluster DMFT for two-dimensional 
Hubbard model at several hole dopings 
(broadening $\gamma=0.02$~eV, $U=2.5$~eV and $T=0.01$~eV). 
(a) $\delta=0.035\ldots0.20$; (b) $\delta=0.31\ldots0.69$.
The dashed lines denote the bare density of states.
The pseudogap is located about 50 meV above $E_F=0$ at low doping up to
$\delta\approx 0.1$. 
The lower edge of the upper Hubbard band shifts from about 
1.5~eV at low doping to 3~eV at large doping. 
}\label{dos}\end{figure}

Figure 1 shows the spectral distributions for a series of hole doping 
concentrations. For simplicity, we show the ED cluster spectra as they 
can be evaluated directly at real $\omega$ without requiring analytical 
continuation. Since we are primarily concerned here with integrated sections 
of the density of states, these cluster spectra are adequate for our analysis. 
The electron density per spin is given by $n_\sigma=0.5(1-\delta)$. 
At all dopings, the upper Hubbard band is seen to be separated from the 
low-energy states by a broad minimum related to the Mott gap in the half-filled 
limit. The lower edge of this band gradually shifts 
from $\omega=1.5$~eV at low doping to about 3.0~eV at large doping.  
The low-energy states close to $E_F=0$ reveal a pseudogap at about 50~meV 
for doping up to about 0.1. 
In Ref. \cite{prb09} we showed that the origin of this pseudogap can be 
traced back to a prominent collective mode in the imaginary part of the 
$(\pi,0)$ component of the self-energy. (See also Ref.\cite{jarrell2001})
This mode is 
therefore directly linked to spatial fluctuations within the $2\times2$
clusters and cannot be described within single-site DMFT. Electron-addition 
states in the vicinity of this collective mode are highly damped, 
giving rise to a pseudogap. In fact, the energy and strength of this 
mode exhibit a clear dispersion with doping, which translates into a 
corresponding doping variation of the mean position and width of the 
pseudogap. Moreover, since the  collective mode is located slightly above 
$E_F$, states in this energy range have a much shorter lifetime than 
states below $E_F$, giving rise to a pronounced particle-hole asymmetry. 
As shown in Ref.\cite{prb09} the low-energy region of the DMFT lattice 
spectra, obtained 
via analytical continuation of the cluster self-energy to real $\omega$, 
are fully compatible with the cluster spectra given here in Fig.~1. These 
lattice spectra are also qualitatively consistent with the corresponding 
low-energy range of the DCA spectra for 16-site clusters (using $t'=0$) 
derived in Ref.\cite{phillips}

\begin{figure} 
\begin{center}
\includegraphics[width=5.5cm,height=6.5cm,angle=-90]{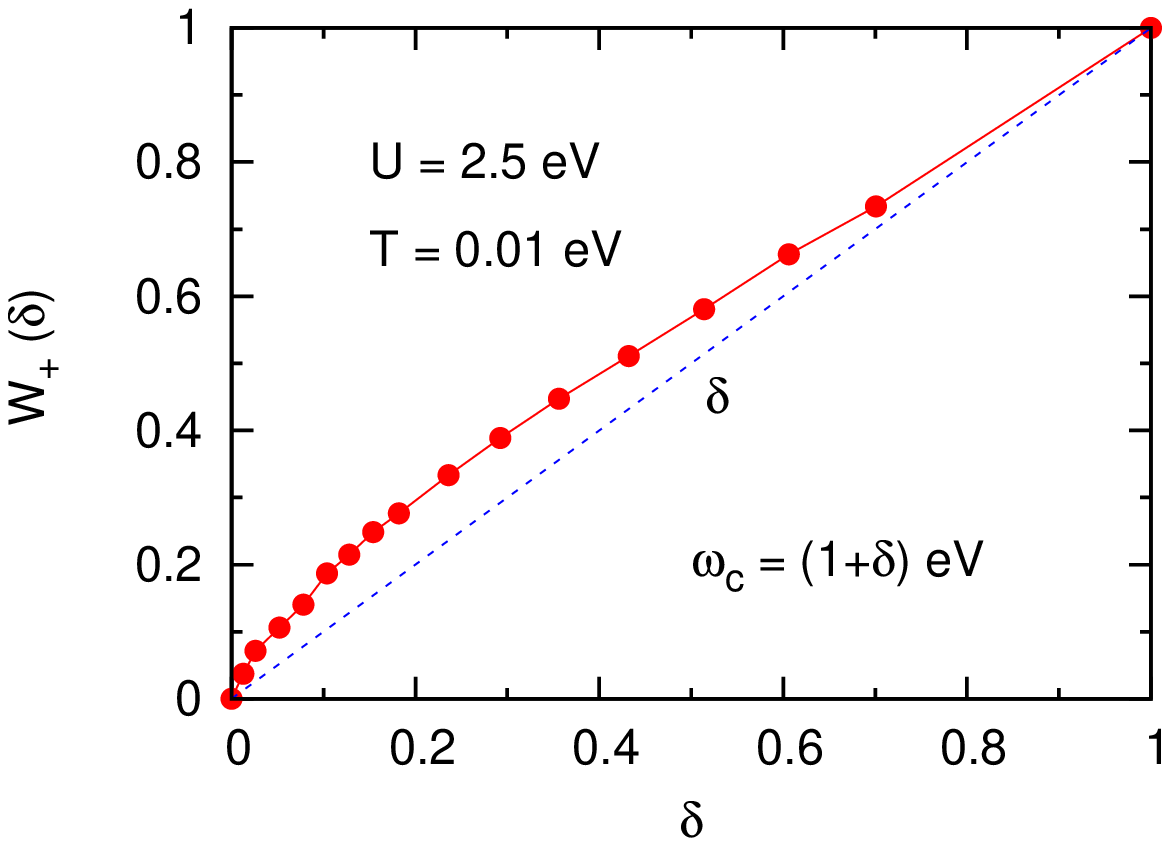}
\end{center}
\vskip-7mm \ \ \ (a)\hfill  \mbox{\hskip5mm}
\begin{center}
\vskip-5mm
\includegraphics[width=5.5cm,height=6.5cm,angle=-90]{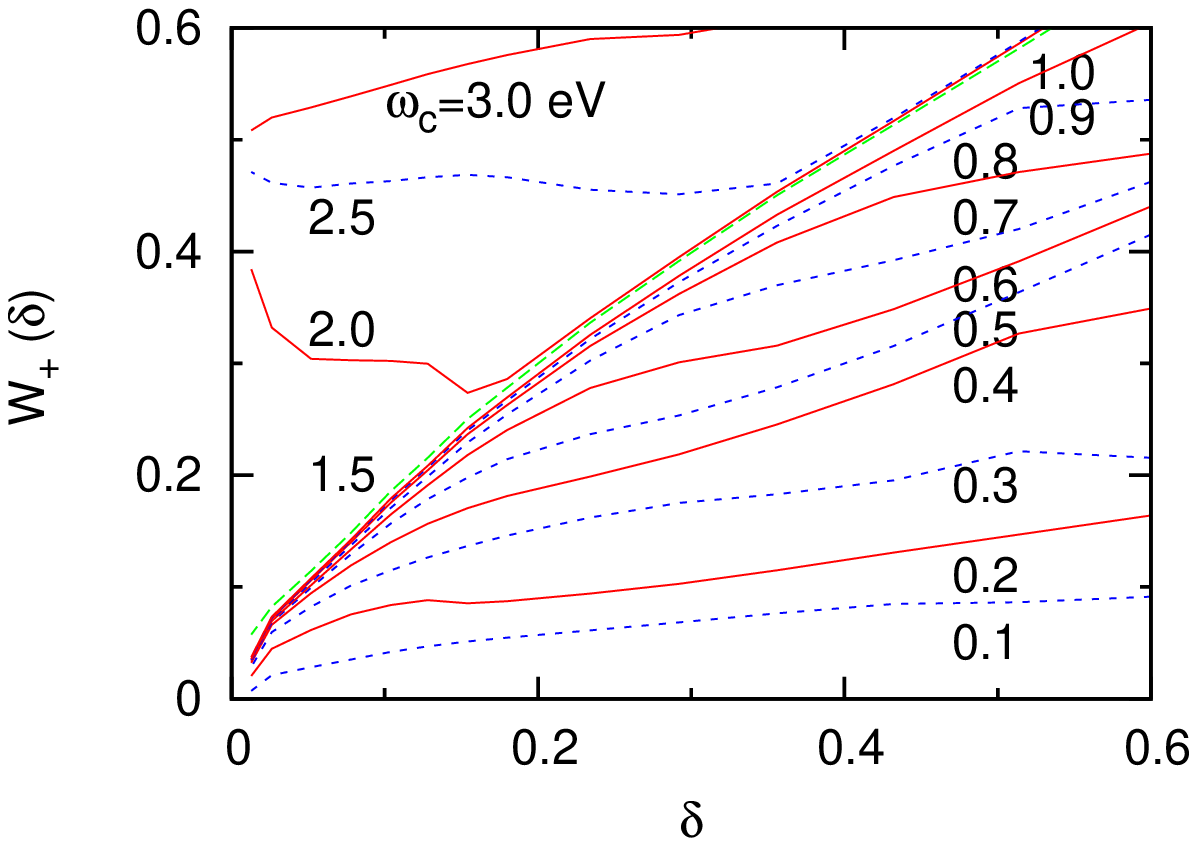}
\end{center}
\vskip-7mm \ \ \ (b)\hfill  \mbox{\hskip5mm}
\caption{(Color online)
Integrated spectral weight of low-energy states per spin as a function of hole 
doping. (a) Integration window extends to minimum below the upper Hubbard band, 
with doping-dependent cutoff energy $\omega_c\approx (1+\delta)$~eV.
(b) Integration window extends up to doping-independent cutoff energy
 $\omega_c= 0.1\ldots 3.0$~eV. 
}\label{weight}\end{figure}

The spectra in Fig.~1 demonstrate that the low-energy states induced
via hole doping are concentrated near $E_F$ only at very low doping. 
With increasing doping, these states spread over a larger energy window,
until at unit doping an uncorrelated empty band appears between 
$E_F$ and the shifted upper band edge at 2~eV. Evidently, upon hole 
doping, spectral weight transfer from the upper Hubbard band is limited 
to the region close to $E_F$ only at very low doping. At larger doping, 
unoccupied spectral weight is generated in the entire range up to $E_F+W$.

Figure 2(a) shows the integrated spectral weight of the doping-induced 
low-energy states, $W_+(\delta)$, where the upper edge  $\omega_c$ of the 
energy window lies in the broad minimum below the upper Hubbard band.
In the present case a convenient choice is  $\omega_c=(1+\delta)$~eV. 
(The low-doping behavior of $W_+(\delta)$ was previously shown in 
Fig.~5 of Ref.\cite{prb09})  
The results demonstrate that $W_+(\delta)\ge \delta$ in the entire doping 
range, consistent with the predictions by Eskes {\it et al.}\cite{eskes}  

Figure 2(b) shows the integrated spectral weight for a variety of fixed, 
doping-independent cutoff energies. For $\omega_c$ in
the range of the pseudogap, i.e., $\omega_c\approx 0.2\ldots 0.3$~eV, 
$W_+(\delta)$ remains much smaller than $\delta$ and approximately 
saturates near $\delta\approx 0.2\ldots0.3$. The reason for this 
saturation is that, as pointed out above, spectral weight 
transfer from the upper Hubbard band must proceed at larger doping 
progressively to states farther above $E_F$, eventually covering the range 
up to $E_F+W$. 
For $\omega_c \approx 1.0\ldots 1.5$~eV, $W_+(\delta)\ge\delta$,
in agreement with Fig.~2(a). For $\omega_c > 1.5$~eV, the energy window
at low doping includes part of the upper Hubbard band (see Fig.~1),
so that  $W_+(\delta)$ does not reach zero in the small doping limit.
In the limit $\omega_c\gg 0$ (not shown), $W_+(\delta)$ includes all
unoccupied states. Thus, $W_+(\delta)=1-n_\sigma= 0.5(1+\delta)$.  
To capture properly the integrated spectral weight of the doping-induced
low-energy states it is evidently necessary to adjust the cutoff energy
so that it roughly tracks the lower edge of the upper Hubbard band.

\begin{figure} 
\begin{center}
\includegraphics[width=10.0cm,height=6.5cm,angle=-90]{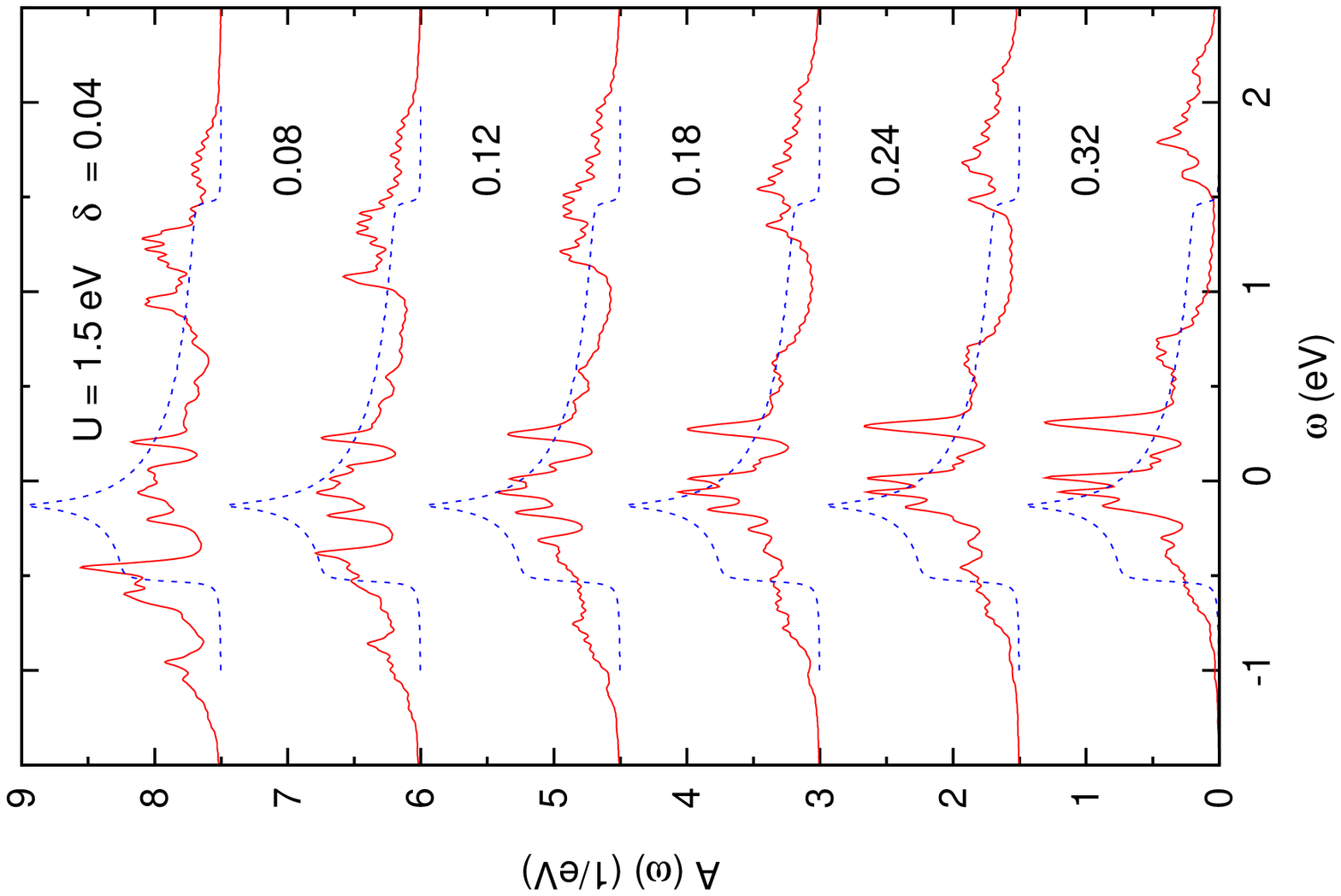}
\end{center}
\vskip-7mm \ \ \ (a)\hfill  \mbox{\hskip5mm}
\begin{center}
\vskip-5mm
\includegraphics[width=5.5cm,height=6.5cm,angle=-90]{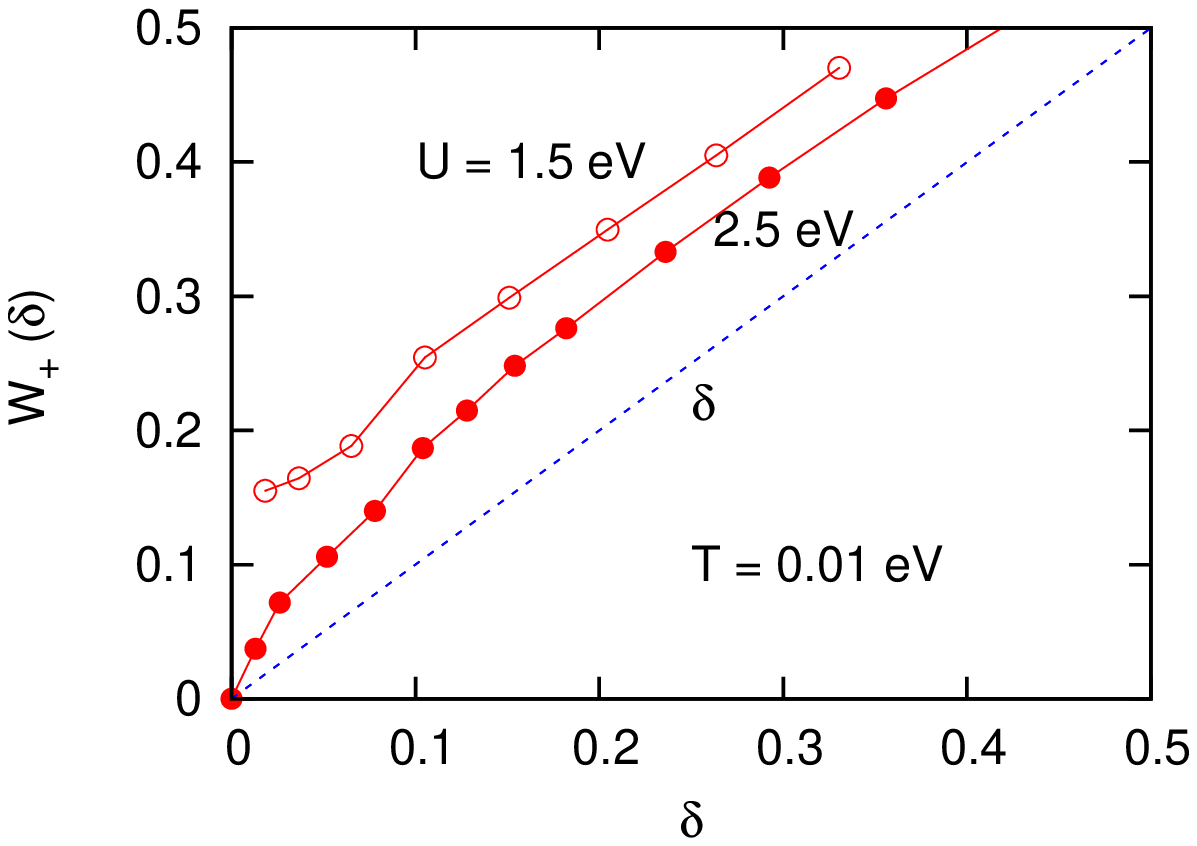}
\end{center}
\vskip-7mm \ \ \ (b)\hfill  \mbox{\hskip5mm}
\caption{(Color online)
(a) Spectral distributions calculated within cluster DMFT for two-dimensional 
Hubbard model at several hole dopings 
(broadening $\gamma=0.02$~eV, $U=1.5$~eV and $T=0.01$~eV). 
The dashed lines denote the bare density of states.
(b) Integrated spectral weight of low-energy states per spin as a function 
of hole doping. The integration window extends to the doping-dependent
minimum below the upper Hubbard band. Empty dots: $U=1.5$~eV; solid dots:
 $U=2.5$~eV. 
}\label{u=1.5}\end{figure}

The cluster DMFT results discussed above are for $U=2.5$~eV. As can be 
seen in Fig.~1, the upper Hubbard band is separated from the itinerant 
unoccupied states by a rather wide gap of about $1\ldots2$~eV, regardless
of doping. The low-energy spectral peak can therefore be defined without
ambiguity. Since there is some uncertainty concerning the appropriate 
value of $U$ for the high-$T_c$ cuprates, we have calculated the spectral 
distributions also for $U=1.5$~eV. As shown in Fig.~3(a), the upper
Hubbard band then lies about 1~eV lower and overlaps the upper part of 
the uncorrelated density of states. The distinction between localized 
and itinerant states is therefore less clear than for $U=2.5$~eV.
Nonntheless, the high-energy peak is still separated from the low-energy 
region by a shallow minimum, except in the limit of small doping. 
At this Coulomb energy, the system is barely insulating at half-filling, 
so that the separation between itinerant and localized states at small 
doping becomes highly ambiguous. 

Figure 3(b) shows the integrated weight of the low-energy feature, where
the minimum close to the lower edge of the upper Hubbard band is used
as cutoff energy, i.e., $\omega_c\approx 0.7\ldots 1.4$~eV for 
$\delta=0.035\ldots0.32$. Because of the uncertainties pointed out above,
$W_+(\delta)$ does not approach zero in the low-doping limit. Nevertheless,
$W_+(\delta)\ge\delta$, in agreement with the prediction by Eskes 
{\it et al.}\cite{eskes} Compared to the analogous spectral peak 
for $U=2.5$~eV, $W_+(\delta)$ is slightly larger for $U=1.5$~eV. This 
trend is also consistent with the results discussed in Ref.\cite{eskes}
The reason for this enhancement is the less clear distinction between 
itinerant and localized unoccupied states for smaller $U$, i.e., some
of the incoherent weight spreads to lower energies. Inspite of these 
uncertainties, it is evident that  $W_+(\delta)$ does not exhibit saturation.       

The experimental spectra by Peets {\it et al.}\cite{peets} reveal
two fairly clearly separated regions: a low-energy feature
and a peak at about $1.0\ldots1.5$~eV higher energy. It 
seems plausible therefore to associate the former with the doping-induced
low-energy states and the latter with the upper Hubbard band, in qualitative
correspondence with the spectra shown in Fig.~1. Since the 
lower peak has an intrinsic width of about 1~eV, it is clearly not 
possible to associate its spectral weight only with the much narrower 
pseudogap region. Instead, this low-energy peak most likely covers
all unoccupied states below the upper Hubbard band. The fact that this 
peak saturates near $\delta\approx 0.2$ in the experiment 
is therefore in conflict with the cluster DMFT results for the single-band 
Hubbard model shown in Fig.~2(a). According to the results provided in 
Fig.~2(b), the saturation obtained by Phillips and Jarrell\cite{phillips} 
appears to be related to the small, doping-independent energy.  

\section{conclusion}

The integrated spectral weight of states induced in the Mott gap
of the two-dimensional single-band Hubbard model via hole doping
is examined within cluster DMFT based on finite-temperature exact
diagonalization. The doping variation of this weight is shown to 
depend sensitively on the energy window in which the low-energy
states are counted. If the cutoff energy is chosen to vary with
doping and to lie in the minimum below the upper Hubbard band, 
qualitative agreement with model predictions by Eskes {\it et al.} 
is found. On the other
hand, if the cutoff energy is taken to be independent of doping,
the integrated spectral weight exhibits a completely different
variation with doping. In particular, at small cutoff energies,
approximate saturation at low doping is found. Since the XAS data
by Peets {\it et al.} require integration over all unoccupied states
except for the upper Hubbard band, we conclude that these data
are not compatible with present cluster DMFT predictions for the 
single-band  Hubbard model. 

\bigskip
I like to thank George Sawatzky for sending the preprint (Ref. 6) 
prior to publication and for numerous discussions. I also like to 
thank Andr\'e-Marie Tremblay for stimulating discussions.


\begin{thebibliography}{99} 

\bibitem{eskes}
    H. Eskes, M. B. J. Meinders, and G. A. Sawatzky, 
       Phys. Rev. Lett. {\bf 67}, 1035   (1991);
    M. B. J. Meinders, H. Eskes, and G. A. Sawatzky, 
       Phys. Rev. B {\bf 48}, 3916 (1993).

\bibitem{peets}
    D. C. Peets, D. G. Hawthorn, K. M. Shen, Y.-J. Kim, D. S. Ellis, H. Zhang,  
    S. Komiya, Y. Ando, G. A. Sawatzky, R. Liang, D. A. Bonn, and W. N. Hardy,
    Phys. Rev. Lett. {\bf 103}, 087402 (2009).

\bibitem{phillips}
    Ph. Phillips and M. Jarrell,
    arXiv:1003.3412 (unpublished). 

\bibitem{hettler}
   M. H. Hettler, A. N. Tahvildar-Zadeh, M. Jarrell, T. Pruschke, and
   H. R. Krishnamurthy, Phys. Rev. B {\bf58}, R7475 (1998).

\bibitem{jarrell2001}
   M. Jarrell, Th. Maier, M. H. Hettler, and A. N. Tahvildarzadeh, 
   Europhys. Lett. {\bf 56}, 563 (2001);
   M. Jarrell, Th. A. Maier, C. Huscroft, and S. Moukouri, 
   Phys. Rev. B {\bf 64}, 195130 (2001).

\bibitem{peets2}
    The same conclusion was reached independently by: 
    D. C. Peets, D. G. Hawthorn, K. M. Shen, G. A. Sawatzky, R. Liang, 
    D. A. Bonn, and W. N. Hardy, arXiv:1004.1146  (unpublished). 

\bibitem{dmft}
   A. Georges, G. Kotliar, W. Krauth and M. J. Rozenberg, 
   Rev. Mod. Phys. {\bf 68}, 13 (1996).  

\bibitem{kotliar01}
   G. Kotliar, S. Y. Savrasov, G. Palsson, and G. Biroli, 
   Phys. Rev. Lett. {\bf 87}, 186401 (2001).

\bibitem{prb09}
   A. Liebsch and N.-H. Tong,
      Phys. Rev. B {\bf 80}, 165126 (2009).

\end{thebibliography}
\end{document}